\documentclass{article}
\usepackage[T1]{fontenc} %
\usepackage[utf8]{inputenc} %
\usepackage{ismir,amsmath,cite,url}
\usepackage{graphicx}
\usepackage{textcomp}
\usepackage{paralist}
\usepackage{multirow}
\usepackage{booktabs}
\usepackage{siunitx}
\usepackage{color}
\usepackage{listings}
\def\noattrib{}  %

\title{Jam-ALT: A Formatting-Aware Lyrics Transcription Benchmark}

\multauthor
{Ondřej Cífka \hspace{0.5cm} 
Constantinos Dimitriou \hspace{0.5cm}
Cheng-i Wang} {
\bfseries{Hendrik Schreiber} \hspace{0.5cm}
\bfseries{Luke Miner} \hspace{0.5cm}
\bfseries{Fabian-Robert Stöter}\\[0.1cm]
AudioShake\\
{\tt\small ondrej@audioshake.ai%
}
}

\def\authorname{O.\ Cífka, C.\ Dimitriou, C.\ Wang, H.\ Schreiber, L.\ Miner, and F.-R.\ Stöter}

\usepackage[bookmarks=false,pdfauthor={\authorname},pdfsubject={\papersubject},hidelinks]{hyperref}
\usepackage[capitalize,nameinlink]{cleveref}

\newcommand{\postbreaksymbol}{\textcolor{blue}{$\hookrightarrow$}}

\newcommand{\punct}{\texttt{P}} %
\newcommand{\paren}{\texttt{B}} %
\newcommand{\nl}{\texttt{L}} %
\newcommand{\sect}{\texttt{S}}
\newcommand{\case}{\texttt{Aa}}

\newcommand{\ctr}[1]{\multicolumn{1}{@{}c@{}}{#1}}
\newcommand{\nan}{\multicolumn{1}{@{}c@{}}{---}}

\begin{document}

\maketitle
\begin{abstract}
Current automatic lyrics transcription (ALT) benchmarks focus exclusively on word content and ignore the finer nuances of written lyrics including formatting and punctuation, which leads to a potential misalignment with the creative products of musicians and songwriters as well as listeners' experiences. For example, line breaks are important in conveying information about rhythm, emotional emphasis, rhyme, and high-level structure.
To address this issue, we introduce \emph{Jam-ALT}, a new lyrics transcription benchmark based on the JamendoLyrics dataset~\cite{durand-2023-contrastive}.
Our contribution is twofold. Firstly, a complete revision of the transcripts, geared specifically towards ALT evaluation by following a newly created annotation guide that unifies the music industry's guidelines, %
covering aspects such as punctuation, line breaks, spelling, background vocals, and non-word sounds. 
Secondly, a suite of evaluation metrics designed, unlike the traditional word error rate, to capture such phenomena.
We hope that the proposed benchmark contributes to the ALT task, enabling more precise and reliable assessments of transcription systems and enhancing the user experience in lyrics applications such as subtitle renderings for live captioning or karaoke.

\end{abstract}
\section{Introduction}\label{sec:introduction}
Recent advances in general-purpose automatic speech recognition (ASR) models pre-trained on large datasets \cite{baevski-2020-wav2vec2,pmlr-v202-radford23a} have enabled automatic lyrics transcription (ALT) with unprecedented accuracy \cite{ou-2022-wav2vec2-alt,zhuo-2023-lyricwhiz}.
However, to the best of our knowledge, public ALT benchmarks ignore letter case, punctuation and formatting (e.g.\ line break placement, parentheses around background vocals). These features are important for producing a high-quality lyrics transcript suitable for distribution within the music industry \cite{apple-lyrics-guidelines,lyricfind-guidelines,musixmatch-lyric-guidelines} (e.g.\ to be displayed on streaming platforms or in karaoke).
While these features were traditionally not part of the output of ASR, this has changed with state-of-the-art systems like Whisper \cite{pmlr-v202-radford23a}, leading to the need for a more comprehensive benchmark.

A dataset adopted by recent works \cite{gupta2020,demirel2021mstrenet,demirel2021low,ou-2022-wav2vec2-alt,zhuo-2023-lyricwhiz} as an ALT test set is JamendoLyrics \cite{stoller2019end}, originally a lyrics alignment benchmark.
Its most recent (``MultiLang'') version \cite{durand-2023-contrastive} contains 4 languages and a diverse set of genres, making it attractive as a testbed for lyrics-related tasks.
However, we found that, in addition to lacking the above features, the lyrics are sometimes inaccurate or incomplete.
While such lyrics may be perfectly acceptable as input for lyrics alignment (and indeed representative of a real-world scenario for that task), they are less suitable as a target for ALT.

To address these issues and help to guide future ALT research, we present the Jam-ALT benchmark, consisting of (1) a revised version of JamendoLyrics MultiLang that follows industry standards for song lyrics transcription and formatting, and %
(2) a set of automated evaluation metrics designed to capture and distinguish different types of errors relevant to (1).
The dataset and the metrics implementation are released online.\footnote{\url{https://audioshake.github.io/jam-alt/}}

\begin{table*}
    \centering
    \renewcommand{\arraystretch}{0.85}
    \setlength\heavyrulewidth{0.15ex}
    \setlength\lightrulewidth{0.1ex}
    \setlength\aboverulesep{0.3ex}
    \setlength\belowrulesep{0.55ex}
    \scalebox{0.97}{%
    \begin{tabular}{@{~}lr@{~}r@{~}r@{~}r@{~}r@{~}rr@{~}r@{~}r@{~}r@{~}r@{~}rr@{~}rr@{~}rr@{~}r@{~}}
    \toprule
     & \multicolumn{6}{c}{All languages} & \multicolumn{6}{c}{English} & \multicolumn{2}{c}{Spanish} & \multicolumn{2}{c}{German} & \multicolumn{2}{c}{French} \\
     \cmidrule(lr){2-7} \cmidrule(rl){8-13} \cmidrule(rl){14-15} \cmidrule(rl){16-17} \cmidrule(l){18-19}
     & WER & $E_\case$ & $F_\punct$ & $F_\paren$ & $F_\nl$ & $F_\sect$ & WER & $E_\case$ & $F_\punct$ & $F_\paren$ & $F_\nl$ & $F_\sect$ & WER & $E_\case$ & WER & $E_\case$ & WER & $E_\case$ \\
    \midrule
    Whisper v2 & 35.7 & 4.5 & 41.7 & \nan & 69.3 & 3.3 & 43.8 & 3.5 & 31.3 & \nan & 63.0 & 11.2 & 25.7 & 6.5 & 45.4 & 5.3 & \bfseries 27.7 & 3.2 \\
    Whisper v2 +sep & 44.0 & 5.3 & 28.0 & \nan & 61.2 & \nan & 32.3 & 5.3 & 39.2 & \nan & 53.8 & \nan & 38.8 & 7.1 & 65.2 & 5.9 & 43.3 & 3.2 \\
    Whisper v3 & 35.5 & 4.3 & 41.6 & \nan & 73.5 & 1.0 & 37.7 & 4.8 & 40.9 & \nan & 71.5 & 2.6 & 28.6 & 5.0 & 40.7 & \bfseries 4.0 & 34.7 & 3.3 \\
    Whisper v3 +sep & 47.9 & 3.8 & 29.0 & \nan & 65.7 & \nan & 43.0 & 4.1 & 23.3 & \nan & 66.8 & \nan & 61.5 & \bfseries 3.6 & 43.5 & 4.4 & 44.9 & 3.2 \\
    LyricWhiz & \nan & \nan & \nan & \nan & \nan & \nan & 24.6 & 3.5 & 34.0 & \nan & 74.0 & 1.4 & \nan & \nan & \nan & \nan & \nan & \nan \\
    AudioShake & \bfseries 26.0 & \bfseries 3.4 & \bfseries 50.5 & \bfseries 29.4 & \bfseries 82.3 & \bfseries 72.1 & \bfseries 22.1 & \bfseries 3.4 & \bfseries 59.0 & \bfseries 32.4 & \bfseries 80.7 & \bfseries 77.4 & \bfseries 22.5 & 4.1 & \bfseries 24.4 & 4.1 & 34.9 & \bfseries 2.0 \\
    \midrule
    JamendoLyrics & 11.1 & 18.5 & \nan & \nan & 93.3 & 85.3 & 14.4 & 15.3 & \nan & \nan & 88.1 & 77.9 & 14.0 & 15.1 & 5.0 & 32.6 & 10.3 & 12.9 \\
    \bottomrule
    \end{tabular}%
    }
    \caption{Benchmark results (all metrics shown as percentages). WER is case-insensitive word error rate, $E_\case$ is case error rate, the rest are F-measures. ``+sep'' indicates vocal separation using HTDemucs. Whisper results are averages over 5 runs with different random seeds, LyricWhiz over 2 runs (transcripts~-- English only~-- kindly provided by authors); our system (AudioShake) is deterministic, hence the results are from a single run. The last row shows metrics computed between the original JamendoLyrics dataset and our revision.
    For full results, see \cref{tab:full-results} in the appendix.}
    \label{tab:results}
\end{table*}

\section{Dataset}
Different sets of guidelines for lyrics transcription and formatting exist within the music industry; we consider guidelines by Apple \cite{apple-lyrics-guidelines}, LyricFind \cite{lyricfind-guidelines}, and Musixmatch \cite{musixmatch-lyric-guidelines}, from which we extracted the following general rules:
\begin{compactenum}
    \item Only transcribe words and vocal sounds audible in the recording; exclude credits, section labels, style markings, non-vocal sounds etc.
    \item Break lyrics up into lines and sections; separate sections by a single blank line.
    \item Include each word, line and section as many times as heard. Do not use shorthands to denote repetitions.
    \item Start each line with a capital letter; respect standard capitalization rules for each language.
    \item Respect standard punctuation rules, but never end a line with a comma or a full stop.
    \item Use standard spelling, including stan\-dard\-ized spell\-ing for slang/contractions where appropriate.
    \item Transcribe background vocals and non-word vocal sounds if they contribute to the content of the song.
    \item Place background vocals in parentheses.
\end{compactenum}
The original JamendoLyrics dataset adheres to rules 1, 3, and 7, partially 2 and 6 (up to some missing diacritics, misspellings, and misplaced line breaks), but lacks punctuation and is lowercase, thus ignoring rules 4, 5, and 8. Moreover, as mentioned above, we found that the lyrics do not always accurately correspond to the audio.
To address these issues, we revised the lyrics in order for them to obey all of the above rules and to match the recordings as closely as possible.
As the above rules are rather unspecific, we created a detailed annotation guide, which is released together with the dataset.
Each lyric file was revised by a single annotator proficient in the language, then reviewed by two other annotators. %
In agreement with the authors of \cite{durand-2023-contrastive}, one of the 20 French songs was removed following the detection of potentially harmful content.

Examples of lyrics before and after revision can be found in \cref{fig:lyrics-ex-1,fig:lyrics-ex-2} in the appendix.

\section{Metrics}
In the following sections, we first discuss the traditional word error rate and then precision and recall measures for punctuation and formatting.
\subsection{Word and Case Error Rates}
\label{sec:wer}
The standard speech recognition metric, \emph{word error rate} (WER), is defined as the edit distance between the \emph{hypothesis} (predicted transcription) and the \emph{reference} (ground-truth transcript), normalized by the length of the reference.
If $D$, $I$, and $S$ are the number of word \emph{deletions}, \emph{insertions}, and \emph{substitutions} respectively, for the minimal sequence of edits needed to turn the reference into the hypothesis, and $H$ is the number of unchanged words (\emph{hits}), then:
\begin{equation}
    \text{WER} = \frac{S+D+I}{S+D+H} = \frac{S+D+I}{N},
\end{equation}
where $N$ is the total number of reference words.

Typically, the hypothesis and the reference are pre-processed to make the metric insensitive to variations in punctuation, letter case, and whitespace, but no single standard pre-processing procedure exists.
In this work, we apply Moses-style \cite{koehn-etal-2007-moses} punctuation normalization and tokenization, then remove all non-word tokens.
Before computing the WER, %
we lowercase each token to make the metric case-insensitive, but also keep track of the token's original form.
To then measure the error in letter case, 
for every \emph{hit} in the minimal edit sequence, we compare the original forms of the hypothesis and the reference token and count an error if they differ. The \emph{case error rate} ($E_\case$) is then computed by dividing the number $S_\case$ of casing errors by the number of words: $E_\case=S_\case/N$.

\subsection{Punctuation and Line Breaks}
\emph{Punctuation restoration}~-- a common ASR post-pro\-cess\-ing step to recover missing punctuation \cite{pais-2022-capitalization}~-- is usually evaluated using precision and recall:
\begin{equation}
    \begin{gathered}
    P = \frac{\text{\small\# correctly predicted symbols}}{\text{\small\# predicted symbols}},\\
    R = \frac{\text{\small\# correctly predicted symbols}}{\text{\small\# expected symbols}}.
    \end{gathered}
\end{equation}
However, computing the numerator requires an alignment between the hypothesis and the reference.
We propose to leverage the same alignment as used in \cref{sec:wer}, but computed on text that includes punctuation and line breaks.

We use the pre-processing from \cref{sec:wer}, but preserve punctuation tokens and, as in \cite{matusov-etal-2019-customizing,karakanta-etal-2020-42}, add special tokens in place of line and section breaks; this leaves us with four token types: word \texttt{W}, punctuation \punct, parenthesis \paren\ (separate due to its distinctive function), line break \nl, and section break \sect.
After computing the alignment between the hypothesis tokens and the reference tokens, we iterate through it in order to count, for each token type $T\in\{\texttt{W},\punct,\paren,\nl,\sect\}$, its number of deletions $D_T$, insertions $I_T$, substitutions $S_T$, and hits $H_T$.
In general, each edit operation is simply attributed to the type of the token affected (e.g.\ the insertion of a punctuation mark counts towards $I_\punct$).
However, a substitution of a token of type $T$ by a token of type $T'\neq T$ is counted as two operations: a deletion of type $T$ (counting towards $D_T$) and an insertion of type $T'$ (counting towards $I_{T'}$).

We can now use these counts to define a precision, recall, and F-1 metric for each token type:
\begin{equation}
    \begin{gathered}
    P_T = \frac{H_T}{H_T+S_T+I_T}, \hskip0.5em\relax R_T = \frac{H_T}{H_T+S_T+D_T},\\
    F_T = \frac{2}{P_T^{-1}+R_T^{-1}}.
    \end{gathered}
\end{equation}

\section{Results and conclusion}
\cref{tab:results} shows the performance of various transcription systems on our benchmark.
We include Whisper \cite{pmlr-v202-radford23a} %
(\texttt{large-v2} and \texttt{large-v3}), optionally with vocal separation using HTDemucs \cite{rouard-2023-htdemucs}; LyricWhiz \cite{zhuo-2023-lyricwhiz} (combining Whisper with ChatGPT \cite{chatgpt}); and our in-house lyrics transcription system.
For Whisper, which does not output line breaks, we use transcription with timestamps and insert line breaks between the timestamped segments.

Interestingly, vocal separation generally degraded the results for Whisper, except for Whisper \texttt{large-v2} on English, where it improved the WER; upon inspection, we find that with separated vocals as input, Whisper often outputs a transcript in the wrong language.
We also observe that \texttt{large-v3} does not necessarily perform better on lyrics than \texttt{large-v2}.

The improvement from LyricWhiz over plain Whisper in terms of WER is clear and even sharper than in \cite{zhuo-2023-lyricwhiz}, and we even see some improvement in terms of line breaks and punctuation.

We also evaluate the original JamendoLyrics dataset itself on our benchmark in order to show how our revision differs from it; the WER of \SI{11.1}{\percent} ($\sim$\SI{14}{\percent} for English and Spanish) attests to the scale of our revisions.

In conclusion, we have proposed Jam-ALT, a new benchmark for ALT, based on the music industry's lyrics guidelines.
Our results bring clarity into how existing systems differ in their performance on different aspects of the task, and we hope that the benchmark will help guide future research on this topic.

\section{Acknowledgment}
We would like to thank Lau\-ra I\-bá\-ñez, Pa\-me\-la Ode, Ma\-thieu Fon\-taine, Clau\-dia Fal\-ler, and Ka\-te\-ři\-na Apo\-lí\-no\-vá for their help with data annotation.

\bibliography{bibliography}

\onecolumn
\appendix

\begin{table*}[ph]
\centering
\begin{tabular}{llr@{~}rr@{~}r@{~}rr@{~}r@{~}rr@{~}r@{~}rr@{~}r@{~}r}
\toprule
& & \multicolumn{2}{c}{Words} & \multicolumn{3}{c}{Punctuation} & \multicolumn{3}{c}{Parentheses} & \multicolumn{3}{c}{Line breaks} & \multicolumn{3}{c}{Section breaks} \\
\cmidrule(lr){3-4} \cmidrule(lr){5-7} \cmidrule(lr){8-10} \cmidrule(lr){11-13} \cmidrule(l){14-16}
Language & System & WER & \ctr{$E_\case$} & \ctr{$P_\punct$} & \ctr{$R_\punct$} & \ctr{$F_\punct$} & \ctr{$P_\paren$} & \ctr{$R_\paren$} & \ctr{$F_\paren$} & \ctr{$P_\nl$} & \ctr{$R_\nl$} & \ctr{$F_\nl$} & \ctr{$P_\sect$} & \ctr{$R_\sect$} & \ctr{$F_\sect$} \\
\midrule
\multirow[t]{6}{*}{All} & Whisper v2 & 35.7 & 4.5 & 42.4 & 40.9 & 41.7 & \nan & 0.0 & \nan & 87.3 & 57.5 & 69.3 & 55.2 & 1.7 & 3.3 \\
& Whisper v2 +sep & 44.0 & 5.3 & 20.0 & 46.4 & 28.0 & \nan & 0.0 & \nan & 74.2 & 52.1 & 61.2 & \nan & 0.0 & \nan \\
& Whisper v3 & 35.5 & 4.3 & 46.4 & 37.7 & 41.6 & \nan & 0.0 & \nan & 76.9 & 70.4 & 73.5 & 37.5 & 0.5 & 1.0 \\
& Whisper v3 +sep & 47.9 & 3.8 & 28.6 & 29.4 & 29.0 & \nan & 0.0 & \nan & 76.4 & 57.7 & 65.7 & \nan & 0.0 & \nan \\
& AudioShake & \bfseries 26.0 & \bfseries 3.4 & \bfseries 47.4 & \bfseries 54.1 & \bfseries 50.5 & \bfseries 37.2 & \bfseries 24.3 & \bfseries 29.4 & \bfseries 87.9 & \bfseries 77.4 & \bfseries 82.3 & \bfseries 78.7 & \bfseries 66.5 & \bfseries 72.1 \\
& JamendoLyrics & 11.1 & 18.5 & \nan & 0.0 & \nan & \nan & 0.0 & \nan & 96.2 & 90.7 & 93.3 & 84.6 & 85.9 & 85.3 \\
\midrule
\multirow[t]{7}{*}{English} & Whisper v2 & 43.8 & 3.5 & 39.8 & 25.8 & 31.3 & \nan & 0.0 & \nan & 81.2 & 51.6 & 63.0 & 52.3 & 6.3 & 11.2 \\
& Whisper v2 +sep & 32.3 & 5.3 & 35.9 & 43.2 & 39.2 & \nan & 0.0 & \nan & 76.0 & 41.7 & 53.8 & \nan & 0.0 & \nan \\
& Whisper v3 & 37.7 & 4.8 & 46.8 & 36.4 & 40.9 & \nan & 0.0 & \nan & 75.5 & 68.0 & 71.5 & 33.3 & 1.4 & 2.6 \\
& Whisper v3 +sep & 43.0 & 4.1 & 25.4 & 21.5 & 23.3 & \nan & 0.0 & \nan & 70.1 & 63.8 & 66.8 & \nan & 0.0 & \nan \\
& LyricWhiz & 24.6 & 3.5 & 49.0 & 26.2 & 34.0 & \nan & 0.0 & \nan & 87.5 & 64.1 & 74.0 & \bfseries 100.0 & 0.3 & 1.4 \\
& AudioShake & \bfseries 22.1 & \bfseries 3.4 & \bfseries 60.3 & \bfseries 57.7 & \bfseries 59.0 & \bfseries 67.4 & \bfseries 21.3 & \bfseries 32.4 & \bfseries 88.6 & \bfseries 74.0 & \bfseries 80.7 & 78.2 & \bfseries 76.6 & \bfseries 77.4 \\
& JamendoLyrics & 14.4 & 15.3 & \nan & 0.0 & \nan & \nan & 0.0 & \nan & 93.6 & 83.3 & 88.1 & 73.6 & 82.8 & 77.9 \\
\midrule
\multirow[t]{6}{*}{Spanish} & Whisper v2 & 25.7 & 6.5 & 48.4 & 51.6 & \bfseries 50.0 & \nan & 0.0 & \nan & \bfseries 86.2 & 61.4 & 71.7 & \bfseries 100.0 & 0.6 & 3.1 \\
& Whisper v2 +sep & 38.8 & 7.1 & 10.8 & 41.9 & 17.2 & \nan & 0.0 & \nan & 76.9 & 44.6 & 56.4 & \nan & 0.0 & \nan \\
& Whisper v3 & 28.6 & 5.0 & \bfseries 54.3 & 34.2 & 41.9 & \nan & 0.0 & \nan & 75.1 & 72.4 & 73.7 & \nan & 0.0 & \nan \\
& Whisper v3 +sep & 61.5 & \bfseries 3.6 & 31.3 & 26.7 & 28.7 & \nan & 0.0 & \nan & 80.3 & 38.9 & 52.4 & \nan & 0.0 & \nan \\
& AudioShake & \bfseries 22.5 & 4.1 & 43.9 & \bfseries 52.4 & 47.8 & \bfseries 53.1 & \bfseries 29.5 & \bfseries 38.0 & 84.7 & \bfseries 80.8 & \bfseries 82.7 & 72.7 & \bfseries 66.7 & \bfseries 69.6 \\
& JamendoLyrics & 14.0 & 15.1 & \nan & 0.0 & \nan & \nan & 0.0 & \nan & 94.3 & 93.1 & 93.7 & 79.0 & 82.1 & 80.5 \\
\midrule
\multirow[t]{6}{*}{German} & Whisper v2 & 45.4 & 5.3 & 29.2 & 57.6 & 38.7 & \nan & 0.0 & \nan & \bfseries 93.3 & 55.8 & 69.9 & \nan & 0.0 & \nan \\
& Whisper v2 +sep & 65.2 & 5.9 & 19.7 & \bfseries 64.6 & 30.2 & \nan & 0.0 & \nan & 66.3 & 68.6 & 67.5 & \nan & 0.0 & \nan \\
& Whisper v3 & 40.7 & \bfseries 4.0 & 33.4 & 53.6 & 41.2 & \nan & 0.0 & \nan & 79.2 & 64.6 & 71.2 & 50.0 & 0.6 & 1.2 \\
& Whisper v3 +sep & 43.5 & 4.4 & 24.5 & 55.7 & 34.0 & \nan & 0.0 & \nan & 84.2 & 62.9 & 72.0 & \nan & 0.0 & \nan \\
& AudioShake & \bfseries 24.4 & 4.1 & \bfseries 40.8 & 59.8 & \bfseries 48.5 & \bfseries 5.2 & \bfseries 17.9 & \bfseries 8.1 & 88.1 & \bfseries 75.3 & \bfseries 81.2 & \bfseries 78.9 & \bfseries 61.6 & \bfseries 69.2 \\
& JamendoLyrics & 5.0 & 32.6 & \nan & 0.0 & \nan & \nan & 0.0 & \nan & 98.7 & 95.8 & 97.2 & 95.9 & 85.4 & 90.3 \\
\midrule
\multirow[t]{6}{*}{French} & Whisper v2 & \bfseries 27.7 & 3.2 & \bfseries 56.0 & 38.8 & 45.8 & \nan & 0.0 & \nan & 89.5 & 62.3 & 73.4 & \bfseries 100.0 & 0.1 & 1.4 \\
& Whisper v2 +sep & 43.3 & 3.2 & 28.7 & 44.7 & 34.9 & \nan & 0.0 & \nan & 83.7 & 54.6 & 66.1 & \nan & 0.0 & \nan \\
& Whisper v3 & 34.7 & 3.3 & 55.9 & 34.2 & 42.4 & \nan & 0.0 & \nan & 78.3 & 77.4 & 77.8 & \nan & 0.0 & \nan \\
& Whisper v3 +sep & 44.9 & 3.2 & 36.2 & 27.1 & 30.9 & \nan & 0.0 & \nan & 74.5 & 65.0 & 69.4 & \nan & 0.0 & \nan \\
& AudioShake & 34.9 & \bfseries 2.0 & 43.3 & \bfseries 48.7 & \bfseries 45.8 & \bfseries 78.8 & \bfseries 28.0 & \bfseries 41.3 & \bfseries 90.5 & \bfseries 79.9 & \bfseries 84.9 & 87.5 & \bfseries 61.9 & \bfseries 72.5 \\
& JamendoLyrics & 10.3 & 12.9 & \nan & 0.0 & \nan & \nan & 0.0 & \nan & 98.4 & 91.3 & 94.7 & 91.4 & 93.9 & 92.6 \\
\bottomrule
\end{tabular}
\caption{Full benchmark results (all metrics shown as percentages). WER is case-insensitive word error rate, $E_\case$ is case error rate, the rest are precisions, recalls, and F-measures. ``+sep'' indicates vocal separation using HTDemucs. Whisper results are averages over 5 runs with different random seeds, LyricWhiz over 2 runs (transcripts~-- English only~-- kindly provided by authors); our system (AudioShake) is deterministic, hence the results are from a single run. The rows labeled ``JamendoLyrics'' show metrics computed between the original JamendoLyrics dataset and our revision.}
    \label{tab:full-results}
\end{table*}

\lstset{
    inputencoding=utf8,
    frame=single,
    basicstyle={\rmfamily\small},
    columns=fullflexible,
    breaklines=true,
    breakatwhitespace=true,
    breakautoindent=true,
    breakindent=0em,
    postbreak=\mbox{\postbreaksymbol\space},
    showlines=true,
    literate=%
    {-}{-}1%
    {é}{{\'e}}{1}%
    {è}{{\`e}}{1}%
    {à}{{\`a}}{1}%
    {ç}{{\c{c}}}{1}%
    {œ}{{\oe}}{1}%
    {ù}{{\`u}}{1}%
    {É}{{\'E}}{1}%
    {È}{{\`E}}{1}%
    {À}{{\`A}}{1}%
    {Ç}{{\c{C}}}{1}%
    {Œ}{{\OE}}{1}%
    {Ê}{{\^E}}{1}%
    {ê}{{\^e}}{1}%
    {î}{{\^i}}{1}%
    {ô}{{\^o}}{1}%
    {û}{{\^u}}{1}%
    {ë}{{\¨{e}}}1
    {û}{{\^{u}}}1
    {â}{{\^{a}}}1
    {Â}{{\^{A}}}1
    {Î}{{\^{I}}}1
}

\begin{figure}
\centering
    \begin{minipage}[t]{0.51\linewidth}
\begin{lstlisting}
people gonna hate let them do it
shine like it ain't nothing to it
damn you a major influence
skate like there ain't nothing doing
live life don't say nothing to them

spectators
side liners
spending days
coming up with sly comments
that's psychotic why try a tarnish such a fly product
why be mad just cause i got hey
i may never know
wave to the haters that put me on the pedestal talk smack
but they really know i'm incredible
unforgettable young blue eyes
the new guy is on schedule
man behind bars and thats minus the federal
stone giant what the hell
could some pebbles do
while you revel in drama im building revenue
tell them you'll get them tomorrow their ain't nothing stressing you
life goes on lifes goes on
you was the shit even before those lights went on
they gonna trash you even if they like your song
people always gonna judge homie right or wrong
\end{lstlisting}
    \end{minipage}
    \hfill
    \begin{minipage}[t]{0.46\linewidth}
\begin{lstlisting}
People gon' hate, let 'em do it (ah)
Shine like it ain't nothin' to it (that's right)
Damn, you a major influence (oh)
Skate like there ain't nothin' doin'
Live life, don't say nothin' to 'em

Spectators, sideliners
Spendin' days comin' up with sly comments
That's psychotic, why tarnish a fly product?
Why be mad just 'cause I got it? Hey
I may never know, wave to the haters
That put me on the pedestal
Talk smack, but they really know I'm incredible
Unforgettable, young blue eyes, the new guy is on schedule
Man behind bars and that's minus the federal
Stone giant, what the hell could some pebbles do
While you revel in drama, I'm buildin' revenue
Tell 'em you'll get 'em tomorrow, there ain't no stressin' you
Life goes on, life goes on
You the shit even before those lights went on
They gon' trash you even if they like your song
People always gon' judge homie right or wrong




\end{lstlisting}
    \end{minipage}
    \caption{An excerpt from \emph{Crowd Pleaser~-- Jason Miller} (license: CC BY-NC-SA). Left: JamendoLyrics, right: Jam-ALT. %
    }
    \label{fig:lyrics-ex-1}
\end{figure}

\begin{figure}
\centering
    \begin{minipage}[t]{0.51\linewidth}
\begin{lstlisting}
y'a pas que tes pas qui m'inspire
qui roule qui se cambre et se penchent
comme un danger qui m'attire

surtout t'arrêtes pas tu sais que tout s'envolerait pour moi
t'es comme un soleil en été le monde tourne autour de toi
le jour la pluie les marais les saisons de chaud ou de froid
les guerres les paix les traités y'a le monde qui tourne et puis toi
y'a pas que tes pas qui m'inspire
belle j'ai vu des démons dans tes hanches
qui roule qui se cambre et se penchent
comme un danger qui m'attire








\end{lstlisting}
    \end{minipage}
    \hfill
    \begin{minipage}[t]{0.46\linewidth}
\begin{lstlisting}
Y a pas que tes pas qui m'inspirent
Qui roulent, qui se cambrent et se penchent
Comme un danger qui m'attire

Surtout t'arrête pas, tu sais
Que tout s'envolerait pour moi
T'es comme un soleil en été
Le monde tourne autour de toi
Le jour, la pluie, les marais
Les saisons de chaud ou de froid
Les guerres, les paix, les traités
Y a le monde qui tourne, et puis toi

Y a pas que tes pas qui m'inspirent
(Y a pas que tes pas qui m'inspirent)
Belle, j'ai vu des démons dans tes hanches
(Belle, j'ai vu des démons dans tes hanches)
Qui roulent, qui se cambrent et se penchent
(Qui roulent, qui se cambrent et se penchent)
Comme un danger qui m'attire
\end{lstlisting}
    \end{minipage}
    \caption{An excerpt from \emph{Pas que tes pas~-- AZUL} (license: CC BY-NC-SA). Left: JamendoLyrics, right: Jam-ALT.}
    \label{fig:lyrics-ex-2}
\end{figure}

\end{document}